\documentclass[prb,twocolumn,showpacs,amsmath,amssymb,superscriptaddress,floatfix]{revtex4}
\usepackage{flushend}
\usepackage{ graphicx } 
\usepackage{ bm } 
\usepackage{ color }

\newcommand{\be}{\begin{equation}}
\newcommand{\ee}{\end{equation}}
\newcommand{\bea}{\begin{eqnarray}}
\newcommand{\eea}{\end{eqnarray}}

\begin{document}
\title{Topological phase and lattice structures in spin chain models}

\author{ Yan-Chao Li }
\author{ Jing Zhang }
\affiliation{ \textit College of Materials Science and
Opto-Electronic Technology, University of Chinese Academy of
Sciences, Beijing, China }
\author{ Hai-Qing Lin }
\affiliation{ \textit
Beijing Computational Science Research Center, Beijing, China
}

\date{ \today }

\begin{abstract}
Finding new topological materials and understanding the physical
essence of topology are crucial problems for researchers. We study
the topological property based on several proposed
Su-Schrieffer-Heeger (SSH) related models. We show that the
topologically non-trivial phase (TNP) in the SSH model can exist in
other model conditions and can even be caused by anisotropy; its
origin has a close relation with the even-bond dimerization. Further
study shows that the TNP is not determined by the periodicity of the
system, but the parity of the two kinds of bonds plays a significant
role for the formation of the TNP. This finding reveals the
topological invariant of the system and might supply potential
extended topological materials for both theoretical and experimental
researches.

\end{abstract}
\pacs{05.30.Rt, 03.67.-a}


\maketitle

\section{Introduction}

Topological phases, with their remarkable properties, have been
intensively studied in recent years~\cite{Hasan2010,Qi2011}. Many
novel quantum phenomena or quantum physical matters, such as the
quantum Hall effect, topological insulators, and topological
superconductors, are related to topological phases. A prominent
feature of a topological phase is the topologically protected edge
states, which are robust against the specific edge shape and
impurity perturbations and could be useful for applications ranging
from spintronics to quantum computations~\cite{Hasan2010,Zhan2017}.

In contrast to conventional phases of matter which are characterized
by symmetry properties and local order parameters, topological
phases are not determined by a specific state, such as the ground
state, but are related to the whole energy spectrum of the system.
They do not belong to the framework of Landau's symmetry breaking
theory and cannot thereby be characterized by local order
parameters~\cite{Sachdev2000,Feng2007,Wen2004}. Therefore, most
studies on topological matters only focus on their novel phenomena
or how to detect or characterize topological phases using edge
states or topological invariants~\cite{Zhan2017,Wang2018}. Few works
study their formation causes and influence factors that are
significant for a comprehensive understand of a topological phase.
This issue in strongly correlated quantum systems still remains as a
challenge.

In addition, the realization of lattice models using ultra-cold
gases and superconducting quantum circuits has attracted great
attention in recent
years~\cite{Wiese2014,Zohar2016,Dalm2016,Tan2018}. Spin chains can
be experimentally simulated through neutral atoms or polar molecules
stored in an optical
lattice~\cite{Simon2011,Struck2011,Yan2013,Hazzard2014}, trapped
ions~\cite{Porras2004,Kim2009,Kim2010,Islam2011}, and NMR
simulator~\cite{Li2014}. As the trapped molecules as an example, the
trapped molecules have two internal states (the lowest rovibrational
state and an excited rotational state) and can be used to modify a
spin-1/2 degree of freedom. Via a microwave field, the molecules can
be coupled and the dipolar interaction will induce spin exchanges
between pairs of molecules. Recently, in the presence of an external
dc electric field, a spin XXZ model was realized experimentally, and
the spin-spin interactions can be tuned with the external electric
field or modified by choosing different pairs of rotational
levels~\cite{Hazzard2014}.

The topological phenomena could also emerge in the synthetic
systems, such as the cold atoms in optical lattices and qubits in
superconducting quantum circuits
~\cite{Atala2013,Jotzu2014,Aidel2015,Flaschner2016,Duca2016,Leder2016,Meier2016,Tan2018}.
Moreover, as these artificial spin lattices could be applied local
interaction change by a local external field modulation, they could
thereby serves as a nature platform for the study of relation
between topology and structure. In addition, spin-based devices have
their advantages because of the absence of scattering due to
conduction electrons. A zero-temperature spin transport has been
studied in a finite spin-1/2 XXZ chain coupled to fermionic leads
with a spin bias voltage\cite{Lange2018}. For such devices defects
and impurity cannot be neglect and will play quite important roles
in the quantum-state property including the topologically
non-trivial phase (TNP) and its applications. Therefore, in this
paper we study a series of related spin chain models to reveal the
relation of the topological quantum phase and different symmetric
interactions structures.

\section{\label{sec:level2} Model and methods}
The Hamiltonian of the Su-Schrieffer-Heeger (SSH) related XXZ spin
model can be written as follows:
\begin{align}\label{eq:1}
H=&-\sum_{j=1}^N\left[(1+\eta)\left(\sigma_{2j-1}^x\sigma_{2j}^x+\sigma_{2j-1}^y\sigma_{2j}^y+\Delta_{1}\sigma_{2j-1}^z\sigma_{2j}^z\right)\right.
\nonumber\\
&+\left.(1-\eta)\left(\sigma_{2j}^x\sigma_{2j+1}^x+\sigma_{2j}^y\sigma_{2j+1}^y+\Delta_{2}\sigma_{2j}^z\sigma_{2j+1}^z\right)\right],
\end{align}
where ${N}$ is the total number of spins, $\eta$ denotes the
dimerization at even or odd bonds (for convenience, we denote the
 $1-\eta$ interacted bonds as $B^-$ and denote those $1+\eta$ interacted bonds as $B^+$), $\Delta_m$ (${m=1,2}$) describe the
anisotropy of the system arising from the spin-spin interaction on
the $XY$ plane and the $Z$ coordinate direction and can be different
for odd and even bonds, and ${\sigma^\alpha}$ (${\alpha=x,y,z}$) are
the Pauli matrices.

When $\Delta_1=\Delta_2=\Delta=0$ and set
$\sigma^x=\sigma^++\sigma^-$ and $\sigma^y=i(\sigma^+-\sigma^-)$,
Eq.~(\ref{eq:1}) becomes to the spin SSH-XX model. Its Hamiltonian
can be written as follows:
\begin{align}\label{eq:2}
H=&-\sum_{j=1}^N
2{\left[(1+\eta)(\sigma_{2j-1}^+\sigma_{2j}^-+\sigma_{2j-1}^-\sigma_{2j}^+)\right.}
\nonumber\\
 &+\left.
(1-\eta)(\sigma_{2j}^+\sigma_{2j+1}^-+\sigma_{2j}^-\sigma_{2j+1}^+)\right],
\end{align}
this Hamiltonian is soluble and its quantum phase transitions (QPTs)
have been studied from the thermodynamics point of
view~\cite{Perk75}. Actually, it has an analogous expression and a
same position of critical point (CP) with the spinless SSH
model~\cite{Wakatsuki2014,Yu2016}. However, the $\sigma$ operators
here are spin operators and satisfy commutate relation, while the
operators in the Hamiltonian of the spinless SSH model are fermion
operators and satisfy anti-commutate relation for the spinless
model. They actually describe different physical nature.

To study the topology, one feasible method is to find topological
invariants, in which the Berry phase has been successfully used as a
detector to reveals the topological nature of the
system\cite{berry1,berry2}. Under the twisted boundary conditions, a
phase of $\phi$ is imposed in the boundary conditions, and the range
of phase values $2\pi$ is discretized into $M$ points, i.e.,
$\phi_1,\phi_2,\cdots,\phi_M$. Then the Berry phase is defined as
\be \gamma = -i \sum_{l=1}^M \ln U(\phi_l) \ee where $U(\phi_l) =
\psi^*(\phi_l) \psi(\phi_{l+1})$ is the link variable~\cite{Yu2016}.
In the topological region the Berry phase is $\pi$ and in the
topological trivial phase it vanishes.

In addition, the entanglement entropy
($E_v$)~\cite{Osterloh2002,Gu2004,Legeza2006} and the maximum of
quantum coherence $\rm {QC}_{max}$~\cite{Li2018} has been
successfully used to detect different QPTs, therefore, the quantum
phase transitions of the systems are also studied by these two
detectors. The detector $\rm QC_{max}$ is defined as follows
\begin{eqnarray}
QC_{max}=\max
I(\rho_{ab},K_{\sigma_n})=\max-\frac{1}{4}Tr[{\rho}_{ab},K_{\sigma_n}]^2.
\label{eq:qcmax}
\end{eqnarray}
where $[...]$ denotes the commutator, $\rho_{ab}$ is the reduced
density matrix for two nearest-neighbour sites $a$ and $b$, and the
$K_{\sigma_n}$ is written as $K_a\bigotimes I_b$~\cite{EPW63}, in
which the Pauli matrices ${\sigma_n}$ with an arbitrary direction
$\vec{n}$ is an observable at site $a$:
\begin{align}
\sigma_{n}= \left(
  \begin{array}{cccc}
    \cos\theta & \sin\theta e^{-i\varphi}  \\
    \sin\theta e^{i\varphi}& -\cos\theta \\
  \end{array}
\right), \label{eq:sigman}
\end{align}
the maximum is taken by traversing $\theta$ and $\varphi$ from $0$
to $2\pi$. In addition, given a known reduced density matrix
$\rho_{ab}$, the von Neumann entropy $E_v(\rho_{ab})
=-\texttt{Tr}\rho_{ab}\texttt{ln}\rho_{ab}$~\cite{Gu2004}.

To calculate the Berry phase and the detectors, we use the exact
diagonalization (ED) techniques as well as the transfer-matrix
renormalization-group technique (TMRG). The TMRG method is based on
a Trotter-Suzuki decomposition of the partition function of a
system, which maps a d-dimensional quantum system to a
d+1-dimensional classical one; therefore it can directly handle
infinite spin chains for an overview, see
Ref~\onlinecite{Xiang1999}. For accuracy, we take m=200 states for
the TMRG calculations. The Trotter- Suzuki error is less than
$1\times10^{-3}$ and the truncation error is smaller than
$1\times10^{-8}$.
\begin{figure}
\includegraphics[width=0.9\columnwidth]{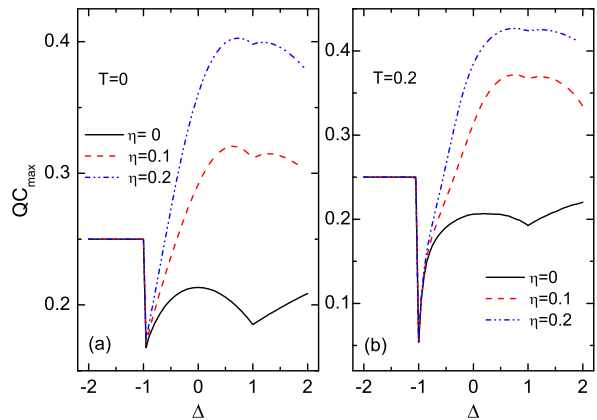}
\caption{\label{fig1} (Color online) $\rm QC_{max}$ under different
$\eta$ from (a) ED method for $N=16$ and (b) TMRG method at
$T=0.2$.}
\end{figure}
\section{\label{sec:level3} QPTs and topologically non-trivial phase in the spin SSH-like systems}\label{sec:3A}
\subsection{Alternated interaction bonds in the spin SSH-XXZ model}

We first study the topological phase and QPTs in the SSH-XXZ model.
When $\eta=0$ and $\Delta_1=\Delta_2=\Delta$, Eq.~(\ref{eq:1})
describes the XXZ spin model. There are two CPs: a continuous phase
transition at $\Delta=1$ and a first-order transition at
$\Delta=-1$~\cite{Takahashi1999}. When $\Delta>1$ it is the
antiferromagnetic (AF) phase with a N\'{e}el order $\left\langle
\sigma_j^z \right\rangle$=-$\left\langle
\sigma_{j+1}^z\right\rangle$, it is the critical Tomonaga-Luttinger
liquid (TLL) phase between $-1<\Delta<1$, and it is the fully
polarized phase at $\Delta<-1$~\cite{Takayoshi2010}. The $\rm
QC_{max}$ results using ED method for $N=16$ are shown in
Fig.~\ref{fig1}(a). The CPs are clearly identified by the two
turning points at $\Delta=\pm1$, which is consistent with the
results in Ref.~\onlinecite{Werlang2010}. When $\eta$ is added, the
two CPs still exist and always keep their positions unchanged. In
other words, the dimerized parameter $\eta$ does not change the the
critical phenomena of the XXZ model. To eliminate the finite-size
effect we modulate the infinite-size case using TMRG numerical
algorithm. The results further confirm the ED results as shown in
Fig.~\ref{fig1}(b).

\begin{figure}[t]
\includegraphics[width=1.0\columnwidth]{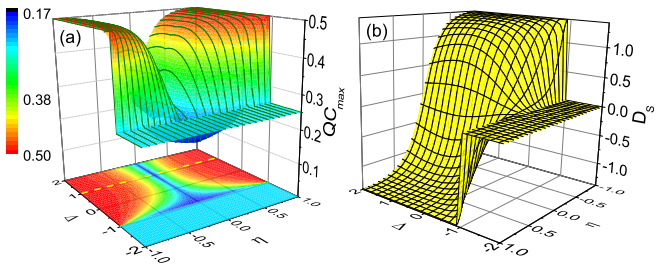}
\caption{\label{fig2} (Color online) (a) $\rm QC_{max}$ and its
contour map and (b) the dimerization $D_s$ as functions of $\Delta$
and $\eta$ for $N=8$. The dashed line in (a) indicates the breaking
point of $\rm QC_{max}$.}
\end{figure}

Figure~\ref{fig2}(a) gives a complete exhibition of the critical
regions. The CPs at $\Delta=\pm1$, which segment the system into the
AF, TLL, and polarized phases, extend to the whole $-1<\eta<1$
region, and there appears a segment at $\eta=0$ for $\Delta>-1$.
Through spin arrangement analysis and the unified structure of $\rm
QC_{max}$, we conclude that $\eta$ does not destroy the AF, TLL, and
polarized orders, but it will introduce the dimerized property.
Figure~\ref{fig2}(b) shows the dimerization $D_s$ defined as
follows~\cite{Legeza2006}:
\begin{align}
D_s=E_{even}-E_{odd} \label{eq:sigman}
\end{align}
where $E_{odd}$ and $E_{even}$ are the neighboring two-site
entropies connected by odd and even bonds in the chain,
respectively. One can see that except the $\Delta<-1$ region and the
$\eta=0$ line, $D_s$ is not zero in the whole parameter regions. The
nonzero $D_s$ confirms the dimerized orders in these regions. Given
the sign of $D_s$, we conclude that, for the AF and TLL phases,
i.e., $\Delta>-1$ region in Fig.~\ref{fig2}(a), there are two kinds
of dimerized phases separated by $\eta=0$ line: it is the even-bond
dimerized structure for $\eta>0$, while it is the odd-bond dimerized
state for $\eta>0$. The polarized phase is not influenced by $\eta$,
it extends to the whole $\Delta<-1$ region.

\begin{figure}
\includegraphics[width=0.7\columnwidth]{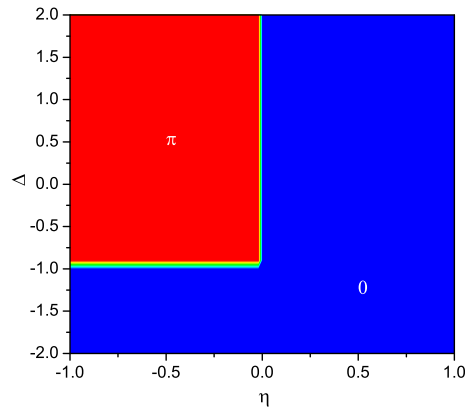}
\caption{\label{fig3} (Color online) Contour map of the Berry phase
as functions of $\Delta$ and $\eta$ under twisted boundary
conditions. The red color $\pi$-value region corresponds to the
topologically non-trivial phase while the blue color $0$-value
region indicates the topologically trivial phase. }
\end{figure}

To further study the topological property, we calculate the twisted
boundary Berry phase as shown in Fig.\ref{fig3}. The topological
nontrivial phase at $\eta<0$ for the spin SSH model extends to a
wide parameter region (see the $\pi$ value in the whole $\Delta>-1$
region) with its CP at $\eta=0$ unchanged. Compared to the dimerized
property $D_s$ in Fig.~\ref{fig2}(b), the topologically non-trivial
phase is consistent with the even-bond dimerized regions. This could
be understood as follows: the topological phase are actually caused
by the competed dimerizations between odd and even bonds. Compared
to the spin SSH model, $\Delta$ term only introduces anisotropy to
the SSH-XXZ model, it does not change the dimerized structure of the
SSH model, therefore, except of the polarized phase which has no
dimerized factors, the topological property also does not change
with $\Delta$.

\subsection{Different dimerized term on the Z direction}
\label{sec:3B}

To further reveal the relations between dimerization and TNP and the
Z-Z interaction effects, we consider $\Delta_1\neq\Delta_2$ in
Eq.~ref{eq:1}, which correspond to the dimerization rate on the Z
direction different from that of the XY plane.

\begin{figure}
\includegraphics[width=1.0\columnwidth]{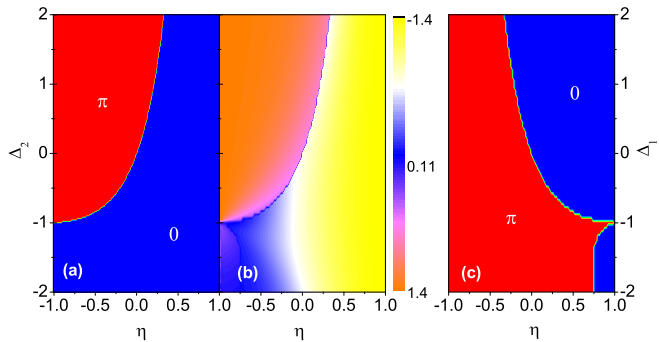}
\caption{\label{fig4} (Color online) Contour map of the twisted
boundary Berry phase as functions of (a) $\Delta_2$ and $\eta$ at
$\Delta_1=0$ and (c) $\Delta_1$ and $\eta$ at $\Delta_2=0$ for
$N=8$; (b) the dimerization $D_s$ at the same condition as (a). }
\end{figure}

For simplify, we consider $\delta_1=0$ and $\delta_2=0$ cases,
respectively. The Berry phase results are shown in
Fig.~\ref{fig4}(a). The $\pi$ value region that indicates the
topological nontrivial phase is quite different with that of the
SSH-XXZ model. When $\Delta_2=0$, the system is the spin SSH model,
the CP is at $\eta=0$. As $\delta_2$ increases, the interaction
along Z direction has a same sign as that of X and Y direction.
Thus, it will enlarge the even-bond dimerizaiton. The regime of the
topological nontrivial phase is then enlarged. When $\delta_2<0$,
the sign of interaction along Z direction is opposite with that of X
and Y direction. It will weaken the antiferromagnetic effect along
the XY direction and will weaken the even-bond dimerized trends. The
results are consistent with this point, the topological nontrivial
phase becomes narrow and narrow and disappears at the critical point
at $\Delta_2=-1$, where the system is polarized along the Z
direction. The dimerization $D_s$ in Fig~\ref{fig4}(b) further
confirms the conclusion. The positive region that corresponds the
even dimerized states consist with the topological nontrivial phase
region in Fig~\ref{fig4}(a).

Just as $\Delta_2$ affect the even-bond dimerized condition,
$\Delta_1$ has a similar effect but on the odd bonds. Thus, it has
an opposite effect for the formation of the even-bond dimerization.
The results are shown in Fig.~\ref{fig4}(c). As $\Delta_1$ change
from positive to negative the region of topological nontrivial phase
is enlarged continuously, which indeed has an opposite effect with
$\Delta_2$. Now we can conclude that the interaction on the Z
direction affects the dimerization of odd and even bonds, and
thereby affects the TNP; The TNP mainly comes from the competition
of the odd and even bond interactions, or more precisely, the
even-bond dimerization plays an important role for the formation of
the TNP.

\subsection{\label{sec:level3c}Topologically non-trivial phase and anisotropy }
The above results show that the topologically non-trivial phase is
mainly caused by the alternated different interactions along the
spin chain. We now consider an uniform interaction between
neighboring bonds but with a different anisotropy conditions between
odd and even bonds. The Hamiltonian is given by
\begin{align}\label{eq:4}
H &= -\sum_j \left[
(1+\beta_1)\left(\sigma_{2j-1}^x\sigma_{2j}^x+\sigma_{2j-1}^y\sigma_{2j}^y\right)\right.
\nonumber \\
&+ \left.(1+\beta_2)
\left(\sigma_{2j}^x\sigma_{2j+1}^x+\sigma_{2j}^y\sigma_{2j+1}^y\right)
\right]
\nonumber \\
&+ \left[ (1-\beta_1)\left(\sigma_{2j-1}^z \sigma_{2j}^z
\right)+(1-\beta_2) \left(\sigma_{2j-1}^z
\sigma_{2j}^z\right)\right],
\end{align}
where $\beta_1$ and $\beta_2$ describe the anisotropy between Z and
XY directions on odd and even bonds, respectively, the operator
$c_{j,\alpha}$ has the same meaning as in Eq.~(\ref{eq:1}).

\begin{figure}
\includegraphics[width=0.7\columnwidth]{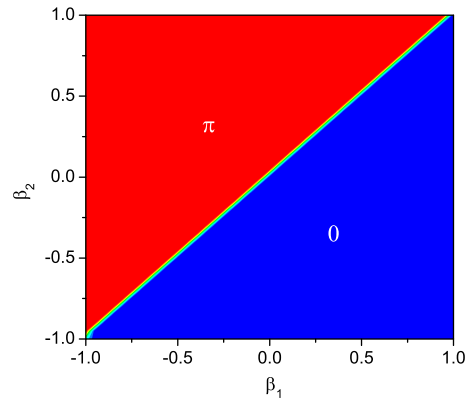}
\caption{\label{fig5} (Color online)  The contour map of twisted
boundary Berry phase as functions of $\beta_1$ and $\beta_2$ for the
anisotropy case at $N=8$. }
\end{figure}

The Berry phase contour map results are shown in Fig.~\ref{fig5}.
Taking the $\beta_1$=$\beta_2$ line as the phase boundary, the
system exhibits the topological nontrivial phase (indicated by the
$\pi$ value in the red region) as long as $\beta_2>\beta_1$. This
condition corresponds to a larger even-bond interaction on the XY
direction. Considering the preceding conclusion that the even-bond
dimerized state is responsible for the formation of the TNP, we
conclude now that the interactions along XY direction is more
dominant than that on the Z direction for the formation of TNP. In
addition, the Berry phase for larger system, such as N=24, shows a
same result. Therefore, the TNP is indeed exist in this anisotropy
case but not caused by the finite-size effects.

\section{\label{sec:level4}Topologically non-trivial phase and lattice structure}
The above results are all related to the difference of interaction
conditions on odd and even bonds, or more precisely the different
interactions between odd and even bonds on the XY direction leads to
the TNP. Then there is an elementary question whether the different
alternated odd and even bonds is a necessary condition for the
formation of the TNP. To answer this question we turn to study the
relationship between the TNP and interactional structures in the
spin chain.

On the other hand, when impurities are considered, the periodically
alternated odd and even bonds structure in the SSH-XXZ model would
be destroyed, and there would be introduced the kinks to the
chain~\cite{Yulu1988}. This condition should be more realistic in
nature. Moreover, the recently developed artificial quantum systems,
such as ultra-cold atom in optical lattices, trapped ions, and
superconducting quantum
circuits, makes it possible to provide tunable and flexible experimental manipulations. 
Therefore, we propose an blocks (marked as ellipse and rectangle,
respectively) alternated structure to describe and simulate the
random arrangement of the two kinds of interactional bonds ($B^+$
and $B^-$).

\begin{figure}
\includegraphics[width=0.9\columnwidth]{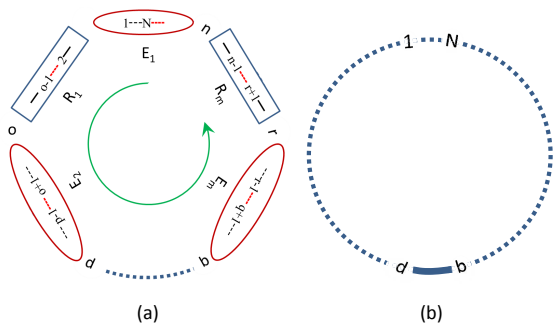}
\caption{\label{fig6} (Color online) Schematic sketch of the random
arrangement of two kinds of interactional bonds. (a)The ellipses and
rectangles circles those continuous arranged $B^-$ and $B^+$ bonds
with coupling constant $1-\eta$ and $1+\eta$, respectively [the
black lines in the ellipses (dashed) and rectangles (solid)
correspond to the $B^-$ and $B^+$ bonds, respectively and the red
dashed lines represent the omitted same bonds in the same block];
the bond number in the $m$-th ellipse and the $m$-th rectangle are
noted as $E_m$ and $R_m$, respectively (to reflect the randomness,
the values of $E_m$ can be different for different $m$. $R_m$ has
the same condition); each lowercase letter represent an arbitrary
spin and the green circular curve with arrow indicates the
increasing direction of the spin arrangement. (b) a limited case
with only one $B^+$ bond between spins p and q while the left bonds
are all $B^-$ bonds. }
\end{figure}

This is sketched in Fig.~\ref{fig6}(a), where the ellipses and
rectangles circles those continuous arranged $B^-$ and $B^+$ bonds
with coupling constant $1-\eta$ and $1+\eta$, respectively; the bond
quantities in the $m$-th ellipse and the $m$-th rectangle are
written as $E_m$ and $R_m$, respectively (for random, the values of
$E_m$ can be different for different $m$, and $R_m$ has a same
condition); the lowercase letters represents different spins and $N$
is the system size. The Halmiltonian can be written as follows:
\begin{align}\label{eq:9}
H=&H_E+H_R \nonumber\\
H_E=&-\sum_{j=1}^{M_E}
(1-\eta)\left(\sigma_{j_l}^x\sigma_{j_r}^x+\sigma_{j_l}^y\sigma_{j_r}^y\right)
\nonumber\\
H_R=&-\sum_{j=1}^{M_R}
(1+\eta)\left(\sigma_{j_l}^x\sigma_{j_r}^x+\sigma_{j_l}^y\sigma_{j_r}^y\right),
\end{align}
where $H_E$ and $H_R$ describe the terms connected by $1-\eta$ and
$1+\eta$ with a total corresponding bond numbers $M_E$ and $M_R$,
respectively, and $j_l$ and $j_r$ are the left and right spin sites
connected by the $j$-th bond, respectively.

When each elliptic block and each rectangular block contain only one
bond, the Hamiltonian describes the spin SSH model as in
Eq.~(\ref{eq:1}) at $\Delta_1=\Delta_2=0$. It becomes the spin XX
model with a defect when there is only one rectangle with one $B^-$
bond while the other bons are all the $B^+$ bonds as sketched in
Fig.~\ref{fig6}(b). The number of the ellipses and rectangles and
the bond number they contained (given by the values of $E_1$ to
$E_m$ and $R_1$ to $R_m$, respectively) determine the bond structure
of the system. We consider a simplest case first, namely, the SSH-XX
bond structure but with some $B^-$ bonds replacing by $B^+$ bonds.
\begin{figure}[b]
\includegraphics[width=1.0\columnwidth]{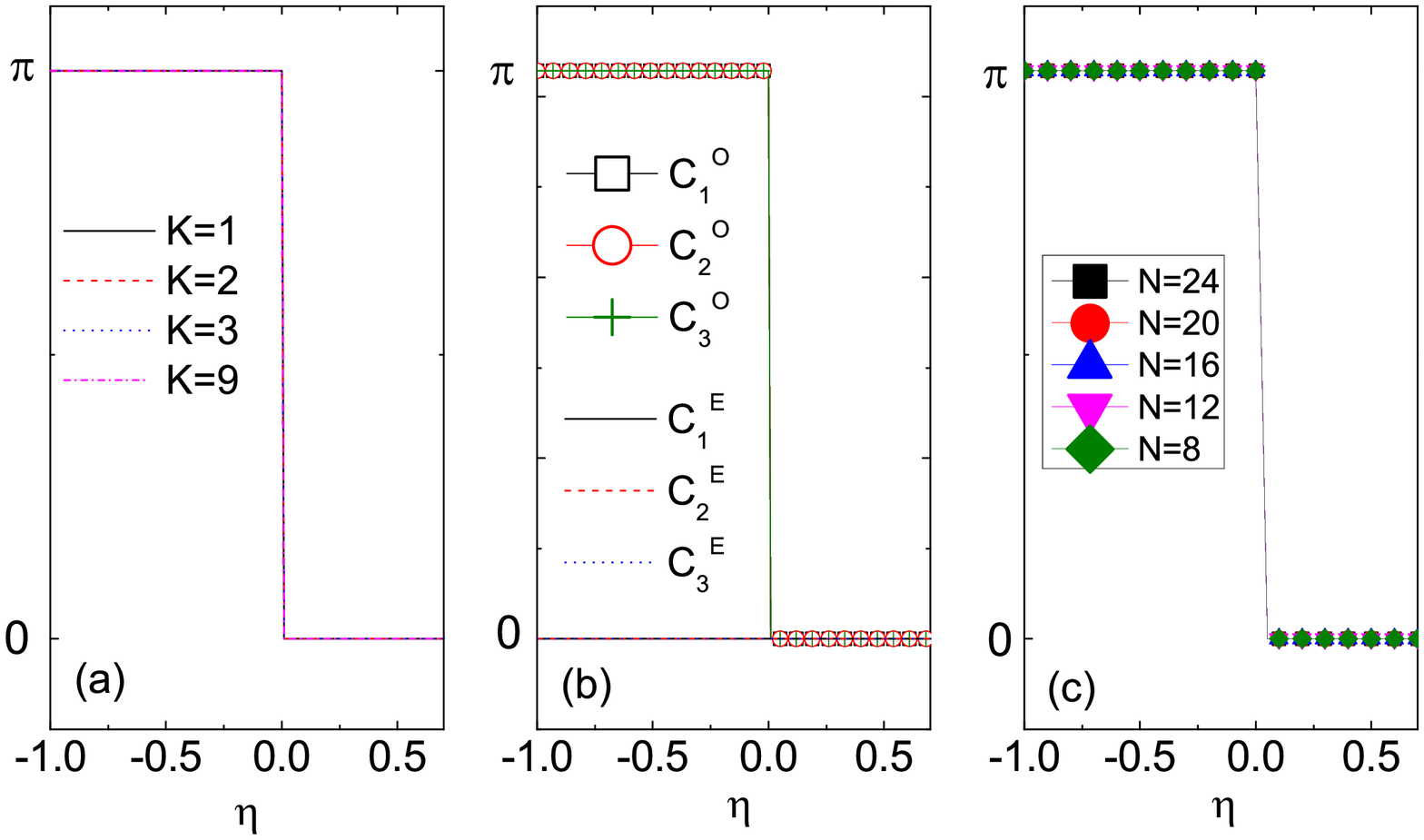}
\caption{\label{fig7} (Color online) Berry phase results for
different spin structures£º (a)the SSH model with $K$ $B^+$ bonds
replaced by $B^-$ bonds£»(b) the line and symbol cures are for $B^+$
and $B^-$ bonds alternated cases: $R_1=R_2=5,E_1=1,E_2=9$ for
$C_1^O$, $R_1=R_2=7,E_1=E_2=3$ for case $C_2^O$, and
$R_1=R_2=3,R_3=7,E_1=E_2=1,E_3=5$ for case $C_2^O$; the lines
without symbol are for the even $B^+$ and $B^-$ bonds alternated
cases: $R_1=R_2=8,E_1=E_2=2$ for $C_1^O$,
$E_1=R_1=E_2=E_3=E_4=R_4=2$ and $R_3=R_5=4$ for $C_2^O$,
$E_1=...E_5=P_1=...=P_5=2$ for $C_3^O$; (c) only $R_1=1$ the left
bonds are all circled by $E_1$ under different $N$.}
\end{figure}
When the number of the replaced bonds is noted as $K$, the Berry
phase as a function of $\eta$ under different $K$ for an $N=20$
chain are plotted in Fig.~\ref{fig7}(a). The $\pi$ value Berry
phase, which indicates the TNP, always exists in the $\eta<0$ region
under different $K$-value cases. This property is consistent with
that of the SSH-XX model. It seems that the changes of bond
structures does not influence the topological property of the
system. Because the periodicity here has been damaged in these
$K\neq0$ cases, the topological property thereby can not be
connected with it. Looking into the structure of these different
cases, there is only one common property: the number of the two kind
bonds ( $B^+$ and $B^-$) at each ellipse and rectangle are odd. This
feature may play a decisive role for the TNP. To further understand
the natural origin of the topology, we turn to the parity of the two
kinds of bonds in each ellipse and rectangle.

We consider various possible odd-numbered $B^+$ and $B^-$ bonds
alternated cases and find they show the same Berry phase feature.
Figure~\ref{fig7}(b) shows the results of three cases of them. The
$\pi$ Berry phase at $\eta<0$ region is completely consistent with
the results in Fig.~\ref{fig7}(a). The TNP indicated by the $\pi$
Berry phase is robust to the bond configurations: as long as the
number of both the $B^+$ and $B^-$ bonds in a continuous block keeps
odd, no matter what the number is. The limiting case sketched in
Fig.~\ref{fig7}(b) satisfies the odd-numbered bonds condition. Thus,
it also hold the same topological property. Figure~\ref{fig7}(c)
shows the results under different system size $N$. The Berry phase
curve does not change as $N$ increases. This feature eliminates the
finite-size effects and confirms the already anticipated topological
property. As the replaced $B^+$ bonds can be regarded as kinks
caused by defects or impurities or can be treated as the local
manipulations in an artificial quantum simulator such as ultra-cold
lattice, this finding points out a class of systems that may have
important potential applications on topology research.

For comparison, we then study the other cases with different bond
parity. The Berry phase results for even-numbered $B^+$ and $B^-$
bonds alternated cases are shown in Fig.~\ref{fig7}(b) as the lines
without symbol. One can see that the Berry phase always equals zero
in all the parameter region. There is no any sign of the TNP.
Afterwards, we further calculate the cases with both odd- and even-
numbered bond blocks at the same time. Although the results show
that there could appear the $\pi$ Berry phase indicated topological
state, the critical points corresponding to the nontrivial to
trivial do not stay at $\eta=0$ any more. Therefore, we conclude
that the topological non-trivial phase in the spin SSH-like systems
comes from the odd-numbered alignment of the two kinds of bonds.

\section{\label{sec:level5} Summary}
In summary, we study the topological property of a class of
one-dimensional spin chains based on several SSH-XXZ related models
by Berry phase and several detectors of QPTs. We considere the TNP
in the spin SSH-XXZ-like model first. Combined with the results of
quantum coherence, entanglement entropy, and Berry phase, we
conclude that the TNP can exist in a wide parameter range, and it
has a quite close relation with the even-bond dimerized state. We
then consider the aspect of anisotropy and found that the TNP can
also appear in an anisotropic system.

To further reveal the physical essence of topology, we study the
relation between the TNP and interactional-bond structures. The
periodicity and parity of the bond arrangements in the chain are
mainly focused. The results show that the TNP is not determined by
the periodicity of the interaction bonds in the system, it can exist
under several kinds of aperiodic cases. On the other hand, the
parity of the two kinds of bonds in their own continuous arranged
part plays a significant role for the formation of TNP. The system
keeps the same topological property as long as it possesses an
odd-odd alternated block arrangement for the two kinds of bonds, no
matter how many blocks it consists and how many bonds are there in
each block, while its topological property will change under other
cases. This finding reveals the topological invariant of the system
and might supply extended topological materials or states by doping
in solid state context or local modulation in quantum-simulator
systems.

%

\begin{acknowledgments}
We acknowledge supports from NSAF U1530401, National Natural Science
Foundation of China under Grant No. 11104009 and computational
resources from the Beijing Computational Science Research Center.
\end{acknowledgments}

\end{document}